\documentstyle[aps,epsfig]{revtex}
\newcommand{\vp}{{\bar p}}
\begin{document}
\twocolumn
\title{Coupled fermion and boson production in a strong background mean-field\vspace*{0.4em}}
\author{D.V. Vinnik, R. Alkofer, S. M. Schmidt}
\address{Institut f\"ur Theoretische Physik, Auf der Morgenstelle 14, Universit\"at T\"ubingen,
D-72076 T\"ubingen, Germany\\[0.6\baselineskip]}
\author{S. A. Smolyansky, V. V. Skokov and A. V. Prozorkevich}
\address{ Physics Department, Saratov State University,
410071 Saratov, Russian Federation \\[0.6\baselineskip]}

\maketitle
\abstract{
We derive  quantum kinetic equations for fermion and boson production
starting from a $\phi^4$ Lagrangian with minimal coupling to fermions.
Decomposing the scalar field into a mean-field part and  fluctuations
we obtain spontaneous pair creation driven by a self-interacting
strong background field.
The produced fermion and boson pairs
  are self-consistently coupled.  Consequently
  back reactions arise from fermion and boson currents determining the
   time dependent self-interacting  background mean-field. We explore
the numerical
  solution in flux tube geometry   for the time evolution of the
mean-field as well as for  the number- and energy densities for
  fermions and bosons. We find
that after a characteristic time all energy is converted from the
background mean-field to particle creation.  Applying this general approach to the production of
``quarks'' and ``gluons'' a typical time scale for the collapse of the
flux tube is $1.5$ fm/c. \\[0.4\baselineskip]
Pacs Numbers: 25.75.Dw,12.38.Mh,05.20.Dd,05.60.Gg}


\section{Introduction}
An ultra-relativistic heavy ion collision is a very complex process
and its theoretical description requires a fundamental understanding
of all phases \cite{QM2000}. After the collision of the nuclei, it is 
expected that
a strongly coupled quark gluon plasma is formed \cite{signals}. The
production of
hard partons due to collisions is one successful way to describe
properties of the pre-equilibrium phase
\cite{Gribovpartons,partonMC,Geiger95,mclerran,Dietrich}. Alternatively the 
flux-tube
model was developed describing the spontaneous creation of soft
particles in a chromo-electric background field \cite{Casher,anders83,stringMC}. 

The idea is based on the QED picture of spontaneous
vacuum pair creation in a strong field \cite{Sauter,Greiner}.
The strong electric background field leads  to a restructuring of the
Dirac sea and drives the system into a false, unstable vacuum
which decays by
emitting particles. Experimentally this was never measured because of
the insufficiently small fields compared to the mass of the electrons
in e.g. optical laser experiments. Planned new 
facilities such as the  X-ray free electron laser (XFEL) may be able to
probe the domain of non-perturbative QED and melt the vacuum \cite{tesla,xfel}. 

In QCD the string
tension is large compared to the mass of the partons. Therefore the
main application of the Schwinger mechanism is for ultra-relativistic
heavy ion collisions to study the 
production of a strongly coupled system of partons within the
flux tube model \cite{Casher}. 
Different versions of a Schwinger-like
source term have been suggested, e.g. \cite{flor,Back,bastirev}, and are implemented in
transport equations to describe formation and evolution of a
quark-gluon plasma on a phenomenological level.

However the precise connection between field theoretical approaches
and a kinetic theory is very challenging and an unsolved problem in general.
For the considered example of  the Dirac/Klein-Gordon equation in an external
field this connection was found in \cite{gsi,kme}. The resulting quantum
Vlasov equation was successfully applied to fermion-  and boson
pair creation \cite{kme,vinnik,eta}. 

In the applications considered so far either
fermions or bosons were treated. One important aspect of most of the
studies was the appearance of plasma oscillations due to back reactions
\cite{Back,oscil} and their damping due to collisions \cite{vinnik,coll}. 
Herein we introduce a
self-consistent scheme in which fermion and boson fields are
coupled via a strong background mean field. This internal field
replaces the external field description of recent approaches and
dynamically couples fermion and boson production.
This coupling is most important for the back reactions since
the total induced current consists of both fermionic and bosonic
contributions.  We  solve the resulting coupled equations in
cylinder geometry \cite{Eisenberg95} to implement physical boundary
conditions leading to a lower bound in the infra-red momentum
region. This procedure not only satisfies the geometry constraints
characteristic for a flux tube but also allows the production of  massless
scalar particles (``gluons''). The fermionic ``quark'' mass  is treated as a
momentum and temperature dependent quantity as obtained from QCD
Dyson-Schwinger equation investigations \cite{bastirev,alk}. 

The article is organized as follows. In Section II we introduce the
effective Lagrangian which we use to derive the  kinetic equations. In
Section III 
we introduce cylindric boundary conditions
for which we solve the equations numerically in Section IV. We
summarize our results in Section V.

\section{Effective Lagrangian and equation of motions}
In this article we will adopt the flux tube picture. One important
aspect is the particular geometry assuming that the ``chromo-electric''
field acts along the longitudinal direction. We introduce the arbitrary
fixed 4-vector 
$n_\mu$ which allows the implementation of the flux tube geometry as
described below.

We start our discussion from the following Lagrangian for fermions
and bosons:
\newpage
\begin{eqnarray}\label{lagrtot}
 L(x)& =& \partial _\mu  \Phi ^* (x)\partial ^\mu \Phi
(x) - m_+^2 \left| {\Phi (x)} \right|^2  - e_+^2 \left| {\Phi (x)}
\right|^4 \nonumber\\
&& + 
ie_+ \left\{ {[\nabla_n \Phi ^* (x)]\Phi ^2 (x) - [\Phi ^* (x)]^2 \nabla_n
\Phi (x)} \right\}\nonumber\\
&& +
\frac{i} {2}\overline \psi  (x)\gamma ^\mu
\stackrel{\leftrightarrow}{\partial} _\mu \psi (x) - m_- \overline
\psi  (x)\psi (x) \nonumber\\
&& -\frac{e_-}{2}\left( \Phi(x)+\Phi^*(x) \right) \overline \psi 
(x)(\gamma n)\psi(x)\,\,,
\end{eqnarray}
where (-) denotes fermions, $\psi$,  and (+) bosons, $\Phi$. The operator
$\nabla_n=n^\mu \partial _{\mu}$ permits to  introduce 
the self-interacting scalar field with first order derivative only.
The total field, $\Phi(x)$, can  be decomposed into a mean field
contribution, $\Phi_0$, and fluctuations around it, $\varphi$, (e.g. \cite{boya})
\begin{equation}\label{80} 
\Phi(x)=\Phi_0(t)+\varphi(x)\,.
\end{equation}
We consider $\Phi_0(t)$ as a neutral, space-homogeneous background
mean field: $\Phi_0=<\Phi>$. The field of
the fluctuations $\varphi(x)$ is in general complex corresponding to a
charged field with a vanishing mean value $<\varphi(x)>=0$.
The Lagrangian (\ref{lagrtot}) contains in particular terms of the order 
$|\Phi(x)|^4$ related to self-interaction. As we will discuss below,
we keep 
the $|\Phi(x)|^4$
contribution to determine {\it only} the background field  $\Phi_0(t)$. That
makes possible a
proper inclusion of back reactions. However we neglect
all higher orders in the fluctuations in deriving the equations of
motion and the
kinetic equations. That means we restrict ourselves to the limit without 
collisions.\footnote{A systematic way of defining the mean-field
approximation and beyond is the large N expansion, see
e.g. \cite{cooper}.} 
We obtain the 
following Lagrangian for the quantum fluctuations in mean field
approximation
\begin{eqnarray}
\label{lagrferm}
L^- (x) &=& \frac{i} {2}\overline \psi  (x)\gamma ^\mu  \stackrel{\leftrightarrow}{\partial}_\mu
\psi (x) - m_- \overline \psi  (x)\psi (x)\nonumber\\ 
&&+e_- \Phi_0(t) \overline \psi(x) (n_\mu \gamma^\mu) \psi (x)\,,\\
\label{lagrbos}
 L^+ (x) &=& \partial _\mu  \varphi^* (x)\partial^\mu  \varphi (x) - m_+ ^2
\left| {\varphi (x)} \right|^2  \nonumber \\
&&+ie_+ \Phi_0(t) \left[ {( \nabla_n \varphi ^* (x))\varphi (x) -
\varphi ^* (x) \nabla_n \varphi (x)} \right]\nonumber\\
&& - e_+^2 |\Phi_0(t)|^2 \left| {\varphi (x)} \right|^2\,.
\end{eqnarray}
The corresponding equations of motion  are found
within standard techniques and read 
\begin{eqnarray}\label{motionferm}
\bigg(i\gamma^\mu\partial_\mu-e_-\gamma^\mu n_\mu 
\Phi^0(t)-m_-\bigg)\psi(x)&=&0\,,\\\label{motionbos}
\left[D^*_\mu D^\mu + m_+^2 \right] \varphi(x)&=&0\,,\\\label{motionmax}
\partial _0^2 \Phi _0(t)  + m_+^2 \Phi _0(t)  + 4 e_+^2 \Phi _0^3(t)- j_-  -
j_+ &=&0\,,
\end{eqnarray}
where Eq. (\ref{motionferm}) is the Dirac equation,
Eq. (\ref{motionbos}) is the Klein-Gordon equation with the covariant
derivative 
\begin{eqnarray}
D_\mu=\partial_\mu + ie_+n_\mu \Phi_0(t)
\end{eqnarray}
and
Eq. (\ref{motionmax}) provides  the time dependent background
mean field. It is important to observe that the external gauge field
which typically appears in applications for pair creation in strong
fields in the Dirac and Klein-Gordon equation 
is replaced by an internal mean field $\Phi _0$. The third term on the
left-hand side of Eq. (\ref{motionmax}) is non-linear and appears as
result of the 
$|\Phi|^4$ self-interacting  contribution in the Lagrangian
(\ref{lagrtot}). The second (mass term) and the third
term  are not contained in the external
field description.\footnote{Note that for a particular
Lagrangian of similar structure, Eq.  (\ref{motionmax}) has a model
specific shape, e.g. starting from a $U_A(1)$ symmetry breaking
Lagrangian of the
Witten-DiVecchia-Veneziano type the non-linear contribution can be  a
transcendent  
function \cite{eta,dahr}.} The equations of motion are self-consistently coupled,
i.e. the mean background field acts on the fluctuations and vice
versa: the produced charged particles generate the currents forming
the background field. 
That is the
well-know back reaction phenomenon. 
The contraction of these currents with  $n^\mu$ reads
\begin{eqnarray}\label{220}
j_- (t) &=&  - e_-  < \overline \psi (x)(\gamma n)\psi (x) >,\\\label{230}
j_+ (t)  &=&  - e_+  < i\varphi ^*
(x)\nabla_n \varphi (x) - i(\nabla_n \varphi ^* (x))\varphi (x)\nonumber\\
&& - 2e_+
\Phi _0 (t)\varphi ^* (x)\varphi (x)  >\,. 
\end{eqnarray}
In the following we will identify
the fermions as ``quarks'' and the bosons as ``gluons'', i.e. we consider the
``gluons'' to be scalar. 
Alternatively one could consider the
bosons as a vector field and such 
an
approach is demonstrated in the Appendix A yielding the same
Lagrangian as Eq.  (\ref{lagrbos}). Many aspects of QCD are not
contained in our approach and clearly to identify the fermionic and
bosonic degrees of freedom with ``quarks and gluons'' is a very optimistic
view. However we hope that some qualitative features discussed below
are robust and also hold for more sophisticated theories than our toy model.

One way to proceed is to solve the equations of motions directly. We
are interested in finding kinetic equations being the exact analogue to
these equations of motion.
Starting from Eqs. (\ref{motionferm}) and  (\ref{motionbos}) it is
possible to derive such  quantum kinetic equations. In \cite{gsi,kme}
it was explained in detail how to introduce quasi-particles,
diagonalize the interaction Hamiltonian and describe the transition
from an  unstable to a stable vacuum as  a dynamical
process employing a
time-dependent Bogolyubov transformation. The resulting equation for the single
particle distribution function  is exact on
the mean field level in a space-homogeneous field, i.e. it preserves the quantum statistical nature and the
pair creation phenomenon.
We choose $n_\mu=(0,0,0,1)$ which is equivalent to considering a
vector potential in temporal gauge with a background field acting in z-direction.
We obtain
\begin{eqnarray}\label{KE}
\frac{{\partial f_\pm (\vp ,t)}} {{\partial t}} +
e_\pm\sigma(t)\frac{{\partial f_\pm (\vp ,t)}} {{\partial p_\parallel }} = S_\pm (\vp,t), 
 \end{eqnarray}
where $\sigma(t)=-d\Phi_0(t)/dt$ is the space-homogeneous time
dependent ``chromo-electric'' field strength, $f_\pm(\vp,t)$ is
the distribution function of the partons in quasi-particle approximation
and $S_\pm(\vp,t)$
is the source term 
\begin{eqnarray} \label{170}
S_\pm(\vp ,t) &=& \frac{1}{2} W_\pm(\vp ,t) \int\limits_{- \infty }^t
{dt'} W_\pm(\vp,t,t')\times\nonumber\\
&& \left[1 \pm 2f_\pm(P_\pm(t,t'),t'
)\right]\cos \left\{ {2 \int\limits_{t' }^t {d \tau} \omega_\pm(\vp;t,\tau )} \right\} 
\end{eqnarray}
describing the creation and annihilation
processes of fermions and bosons  as result of vacuum tunneling
in a strong quasi-classical mean-field $\Phi_0(t)$. 
The transition amplitudes are

\begin{equation}\label{180} 
W_\pm (\vp ;t,t' ) = \frac{{e_\pm \sigma(t' )P_\pm (t,t' )}}
{{\omega _\pm^2 (\vp ;t,t')}} \left[ {\frac{\varepsilon_{\pm\perp} }
{{P_\pm (t,t')}}} \right]^{2s_\pm}\,. 
\end{equation}
For fermions $s_-=1/2$ and for bosons $s_+=0$.
The energy squared of the quasi-particles reads
\begin{eqnarray}\label{190} 
\omega^2_\pm (\vp,t,t')& =& \varepsilon_{\pm \perp }^2  +
P_\pm^2(t,t')\,, \\
P_\pm (t,t')& =& p_\parallel  - e_\pm \int\limits_{t' }^t {d\tau \sigma(\tau )}\,, 
\end{eqnarray}
where $\varepsilon^2_{\pm \perp} = m_\pm^2  + p_1^2  + p_2^2 $ is the
transverse energy. It is assumed that $\mathop {\lim }\limits_{t\to -
\infty } \sigma(t) = 0$ and we simplify the notation by introducing 
\begin{equation}\label{200} 
\omega_\pm (\vp) =  \omega_\pm (\vp,t,t)\,\,, \qquad W_\pm (\vp,t) =  
W_\pm (\vp,t,t)\,. 
\end{equation}
The kinetic equations (\ref{KE}) coincide with recent results and
retain all physical information contained in the field equations (\ref{motionferm})-(\ref{motionbos}). The main
properties of the source term are the inclusion of quantum
statistical effects due to the Fermi suppression (Bose enhancement)
factor for fermions (bosons) and  the nonlocal time structure. 
``Quarks and gluons'' are produced with
a non-trivial, non-equilibrium momentum distribution which is different
in transverse- and in longitudinal direction to which the field
$\Phi_0$ couples according to the chosen anisotropy. Apart from the
conceptual interest of a kinetic formulation, the practical merits are 
obvious: e.g.
the  straightforward identification of the Markovian- or low
density limit and the phenomenologically simple  inclusion of
collisions. 

The background field $\Phi_0$ decays through particle creation and
generation of currents which are readily obtained:
\begin{eqnarray} \label{250}
&&j_\pm(t)=2\,e_\pm g_\pm \int\, \frac{d^{3}p}
{(2\,\pi)^{3}}\bigg(\frac{p_\parallel}{\omega_\pm(\vp)}f_\pm(\vp,t)+\nonumber\\&&\frac{\omega_\pm(\vp)}{e_\pm\sigma(t)}\frac{d
f_\pm(\vp,t)}{dt} - \frac{e_\pm \dot \sigma(t)p_\parallel}{8\omega^4_\pm(\vp)}
\bigg[\frac{\epsilon_{\bot\pm}}{p_\parallel}\bigg]^{2s_\pm}\bigg)\,,
\end{eqnarray}
where $g_\pm$ is the degeneracy factor
The first term is proportional to the occupation number itself and the
second term is related to polarization being proportional to the rate.
The last term  is
a regularizing counter term. Its origin is associated with charge
regularization; further details can be found in \cite{Back}.

\section{Cylindric boundary and the flux tube geometry}
Another element that we would like to introduce in our model is a  
cylindrically
confined region for the  particle production process. A similar
approach was explored in \cite{Eisenberg95,PCFT} and an advanced 
description of a dynamical flux tube was given in
\cite{lampert}. 
The background field is
constructed to act only inside a cylindric tube and is assumed to
vanish outside this  finite volume. We introduce this feature because of
 two reasons: (i) to implement the given anisotropic symmetry
and (ii) to avoid  difficulties with massless particle
production for ``gluons''. It is clear that a finite size in transverse direction leads 
to a quantization of the transverse momentum, so that at zero mass
$m_+=0$ the minimal energy at which particles can be produced is
$\Lambda^{IR}$.

One possible, simple way is to introduce the boundary conditions 
\begin{eqnarray}\label{260}
\Phi(x_3,\rho_0,\phi,t)=0
\end{eqnarray}
to require
that the field vanishes outside a fixed flux tube of the radius
$\rho_0$, e.g. \cite{Eisenberg95}.
The Klein-Gordon equation in such cylinder coordinates reads
\begin{eqnarray}\label{280}
\bigg[\partial_0^2 - \nabla_{\perp}^2 - \partial_3^2 & -&
2ie_+\Phi_0(t) \partial_3-e_+^2\Phi_0^2(t)\nonumber\\
&+& m_+^2 \bigg] \varphi(x_3,\rho,\phi,t)=0,
\end{eqnarray}
where the Laplace operator is given by
\begin{eqnarray}\label{290}
\nabla^2_{\perp}=\frac{\partial^2}{\partial \rho^2}+\frac{1}{\rho}\frac{\partial}{\partial \rho}
+\frac{1}{\rho^2}\frac{\partial^2}{\partial \phi^2}\,.
\end{eqnarray}
The radius  is denoted with $\rho$ and the angle is
$\phi$. Here we assume that the mean field depends only on time and
not explicitly on the radius: $\Phi_0=\Phi_0(t)$.\footnote{ A $\rho$-
dependent field $\Phi_0$ would lead 
to nontrivial changes in the quasi-particle representation itself
and to more complicated final expressions for the
equations of motion.} This assumption allows to make a separable ansatz
of the following form
\begin{eqnarray}
\label{300}
\varphi(x_3,\rho,\phi,t)=N e^{il\phi}e^{-ik_3x_3} T(t)R(\rho)\,.
\end{eqnarray}
Substituting (\ref{300}) into (\ref{280}) we find two equations, one for
the time dependent part 
\begin{eqnarray}\label{310}
\ddot T(t) + \left[ m_+^2+ \left(k_3 - e_+ \Phi_0(t) \right)^2 +\varepsilon^2 \right]T(t)=0
\end{eqnarray}
and one for the spatial dependence
\begin{eqnarray}\label{320}
\rho^2 R^{''} (\rho) + \rho R^{'}(\rho)+\left(\varepsilon^2 \rho^2 - l^2 \right)R(\rho)=0\,,
\end{eqnarray}
where $\varepsilon$ is any constant value.
The latter equation is known as the Bessel equation and with the 
boundary condition $R(\varepsilon \rho_0)=0$. Eq. (\ref{300})
can be rewritten in terms of Bessel functions
\begin{eqnarray}\label{340}
\varphi_{nl,k_3}(x_3,\rho,\phi,t)=e^{-ik_3x_3} T_{nl}(t)
\frac{J_l(\varepsilon_{nl} \rho)}{\sqrt{\pi}
\rho_0 J^{'}_l(\varepsilon_{nl} \rho_0)} e^{il\phi},
\end{eqnarray}
where $\varepsilon_{nl} \rho_0$ is the $n-th$ zero of the Bessel
function $J_l$.
\begin{table}
\begin{center}
\begin{tabular}{c|c|c|c}
$\rho_0$ &1 fm &2 fm & 3 fm\\\hline
 $\Lambda^{IR}$& 0.48 GeV&0.24 GeV&0.16 GeV
\end{tabular}

\vspace{0.2cm}

\caption{The radius of the flux tube $\rho_0$ is fixed and leads to a
discretized momentum with a lower bound $\Lambda^{IR}$.  
\label{table}} 
\end{center}
\end{table}
The resulting kinetic equations are identical to Eqs. (\ref{KE}) but the quasi-particle
energy is now discrete and reads
\begin{eqnarray}\label{350}
\omega_{\pm nl}^2(\vp,t)&=&\varepsilon _{+ \bot nl }^2 + (k_3-e_\pm \Phi_0(t))^2\,,\\
\varepsilon^2_{\pm \bot nl }& =& m_\pm^2  + \varepsilon^2_{nl}\,.
\end{eqnarray}
The kinetic momentum $p_\parallel = k_3-e_\pm \Phi_0(t)$ contains the
time dependent mean-field and the canonical momentum $k_3$. The lowest
possible values for the transverse energy are assumed for  $\varepsilon_{10}$ and the
corresponding  $\Lambda^{IR}=\varepsilon_{10}$ are given in Table I for three
different flux tube radii.
This means that the transverse energy never vanishes. Even for a zero
gluon mass the discretization provides a lower bound for the  momentum.
To proceed we consider all
observable values as mean values in the following way 
\begin{equation}
<<...>>_\rho =\frac{1}{\pi \rho_0^2}\int_0^{\rho_0}\rho d\rho\int_0^{2\pi} d\phi <...>\,. 
\end{equation}
Assuming such mean ($\rho$- independent) quantities is a very simple
approximation. However for the qualitative study performed it is
sufficient and can certainly be improved for a
particular experimental application. The charges in strong coupling
limit are fixed to be $e_\pm = 1$
throughout the numerical calculations and the degeneracy factors for a
three flavor system are for
quarks $g_-= 18 $ and for gluons $g_+= 16$. Quantitatively the
numerical results depend on these values, qualitatively the main
effects discussed below are robust against small changes of all
parameter values.
\begin{figure}[t]
\centerline{\epsfig{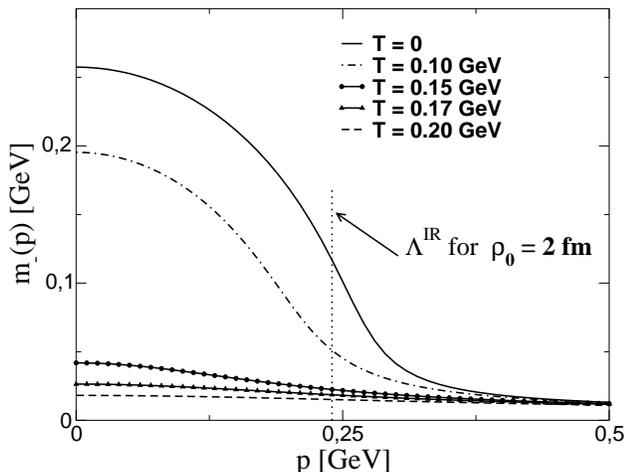}}

\caption{Momentum dependence of the quark mass for different
temperatures obtained from solving the QCD Dyson-Schwinger equation for
the quarks employing a simplified gluon propagator. The fixed
radius of the flux tube, $\rho = 2$ fm, leads to a lower momentum
bound of $\Lambda^{IR}=0.24$ GeV. }
\label{fig1}
\end{figure}
Before we present our numerical results we briefly discuss the mass
scales of the produced particles.
The
``gluon'' mass is chosen to be zero or finite. This is possible
because of the geometry constraint discussed above. The fermion mass is
very small on the hadronic scale in the perturbative regime,
$m_-(p\rightarrow \infty) << \Lambda_{QCD}$. In the
non-perturbative domain dynamical chiral symmetry breaking appears and
the quark mass increases by about two orders of magnitude
$m_-(p\rightarrow 0)\sim\Lambda_{QCD}$. For very
large temperatures the  quarks can be well approximated by their
current quark mass for all momenta. However in vicinity of the phase
transition a dynamical mass is generated. Furthermore,
non-perturbative effects
are manifest up to $ 2 - 3 \times T_c \sim 300 -  450$ MeV.
It was shown in a variety of model approaches that within the parton 
creation model employed in our investigation such temperatures are initially
reached. Due to a rapid expansion the temperature will further
decrease and non-perturbative dressing of the quarks appears. 

In QCD the momentum dependence of the quark mass can be calculated on
the lattice, e.g. in \cite{jonivar} in Landau gauge. QCD
Dyson-Schwinger equation models 
\cite{bastirev,alk,DSE}
are in good agreement with these studies and we  employ a simple
model introduced in \cite{MN} extended to non-zero
temperature and density in \cite{bastirev,TD}. The instantaneous
version of such
an infrared dominant model \cite{bastirev,hirsch} leads to the
following  QCD gap equation for the scalar part of the self-energy:
\begin{eqnarray}\label{mq}
b(\vp)&=&m_0+\eta\frac{m_-(\vp)}{\sqrt{\vp^2+m_-^2(\vp)}}\times\\\nonumber
&&[1-f^{eq}_-(\vp,T,\mu) - {\bar f}^{eq}_-(\vp,T,\mu)]
\end{eqnarray}
and the vector part
\begin{equation}
a(\vp)=\frac{2b(\vp)}{m_0+b(\vp)}
\end{equation}
where $m_-=b/a$. $f^{eq}_-$ and ${\bar f}^{eq}_-$  are the Fermi
distribution functions for particles and antiparticles, respectively.
The value of the mass scale parameter $\eta= 1$ GeV is inspired by the
potential energy in a QCD string at the confinement distance,
$V_{q{\bar q}}(r = 1 {\rm fm}) \sim 1$ GeV. The momentum dependence
calculated with Eq. (\ref{mq}) is depicted in Fig. 1 for different
temperatures. The chemical potential is  $\mu = 0$ and the current
 quark mass $m_0= 5$ MeV. For further calculations we will use the shape
corresponding to $T=T_c\sim 170$ MeV. For the current application a
constant fermion mass with the value of $m_0$ would provide a good guide. However  for future
applications, it will be very useful to have its complete temperature
and momentum dependence introduced already.  

\section{Numerical results}
The final equations solved numerically are given in the Appendix B.
In the following we  compare the  model cases (i)
coupled fermion and boson production with non-linear self-interaction
included; (ii) only boson production with self-interaction, i.e
neglecting the fermionic contribution, $j_-$ in Eq. (\ref{motionmax}) to (iii) boson
production without self-interaction, i.e. neglecting the nonlinear
term in  Eq. (\ref{motionmax}) and $j_-$ . Furthermore we will explore the
dependence of the numerical results on the parameters of the model,
i.e. the masses and the flux tube radius, see Table I.  We  use  (a) a
zero ``gluonic''
mass $m_+ = 0$ and compare the
results with  (b) a finite ``gluon'' mass of $m_+ = 0.5$ GeV; the mass of the
quarks is given by the simple model defined by Eq. (\ref{mq}). 
\subsection{The mean background field}
\begin{figure}[t]
\centerline{\epsfig{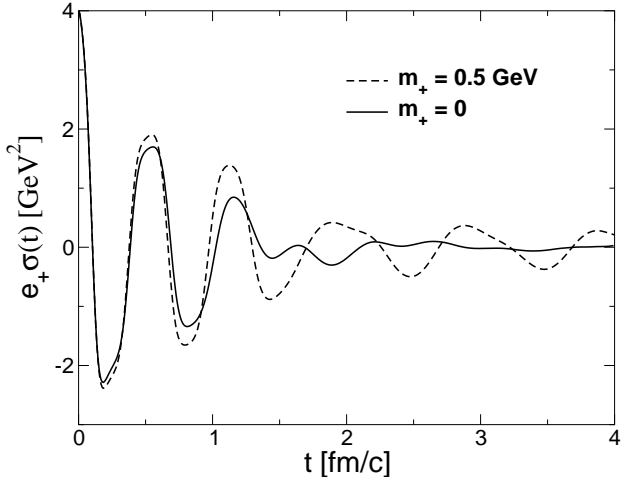}}
\caption{The time dependence of the strong background ``chromo-electric'' field,
$\sigma(t)$, for ``quarks and gluons'' with self-interaction.  The 
flux tube radius is $\rho_0 = 2$ fm and  the solutions for two different ``gluon'' masses
$m_+ = 0$ and $m_+ = 0.5$ GeV are compared . For zero ``gluon'' mass the flux tube
collapses at $\tau_f \sim 1.5 - 2$ fm/c and all energy is transformed from
the mean field to the created particles. For massive ``gluons'' weak plasma
 oscillations survive. 
}
\label{fig2}
\end{figure}

We start the discussion of the numerical results with the background
mean field $\sigma(t)$. As initial condition for the field we choose 
$ e_+\sigma(t=0)= 4$ GeV$^2$ corresponding to a
large initial
energy density of about $\epsilon(t=0)\sim 700$ GeV/fm$^3$ related to
future LHC experiments. 
The time dependence of the ``chromo-electric'' field strength $\sigma(t)$ is plotted Fig. \ref{fig2}.
The self-interaction leads to oscillations on a time
scale of $\tau_{pl}\sim 0.5$ fm/c in the early phase of the
evolution. This result  is basically independent of the masses
used, however it strongly depends on the initial field strength.  
The amplitude of the oscillations is damped due to back reactions and
the background mean-field 
vanishes after a few periods at about $\tau_{f} \sim 3 \tau_{pl}
\sim 1.5$ fm/c for the case of zero gluonic mass. 
With other words: the field leading to pair creation disappears, the flux
tube dynamically collapses.   For  $m_+ = 0.5$ GeV the oscillations
at the beginning of the evolution
are damped as well but at $\tau_f$ the field does not disappear completely. It rather
evolves into a permanently oscillating field with an amplitude of
10\% of the initial value.\footnote{ We suppose that a detailed study
of this effect in the application presented in  \cite{eta,dahr} could
lead to a similar result.} 
This behavior for $t > \tau_f$ can be compared to  typical
results in an external field approach where a constant background field or an
impulse shape was applied and undamped plasma oscillations appeared.
The production of
quarks is much less effected, see Fig \ref{fig3}, and 
a physical explanation of this effect seems to be straightforward:
\begin{figure}[t]
\centerline{\epsfig{figure=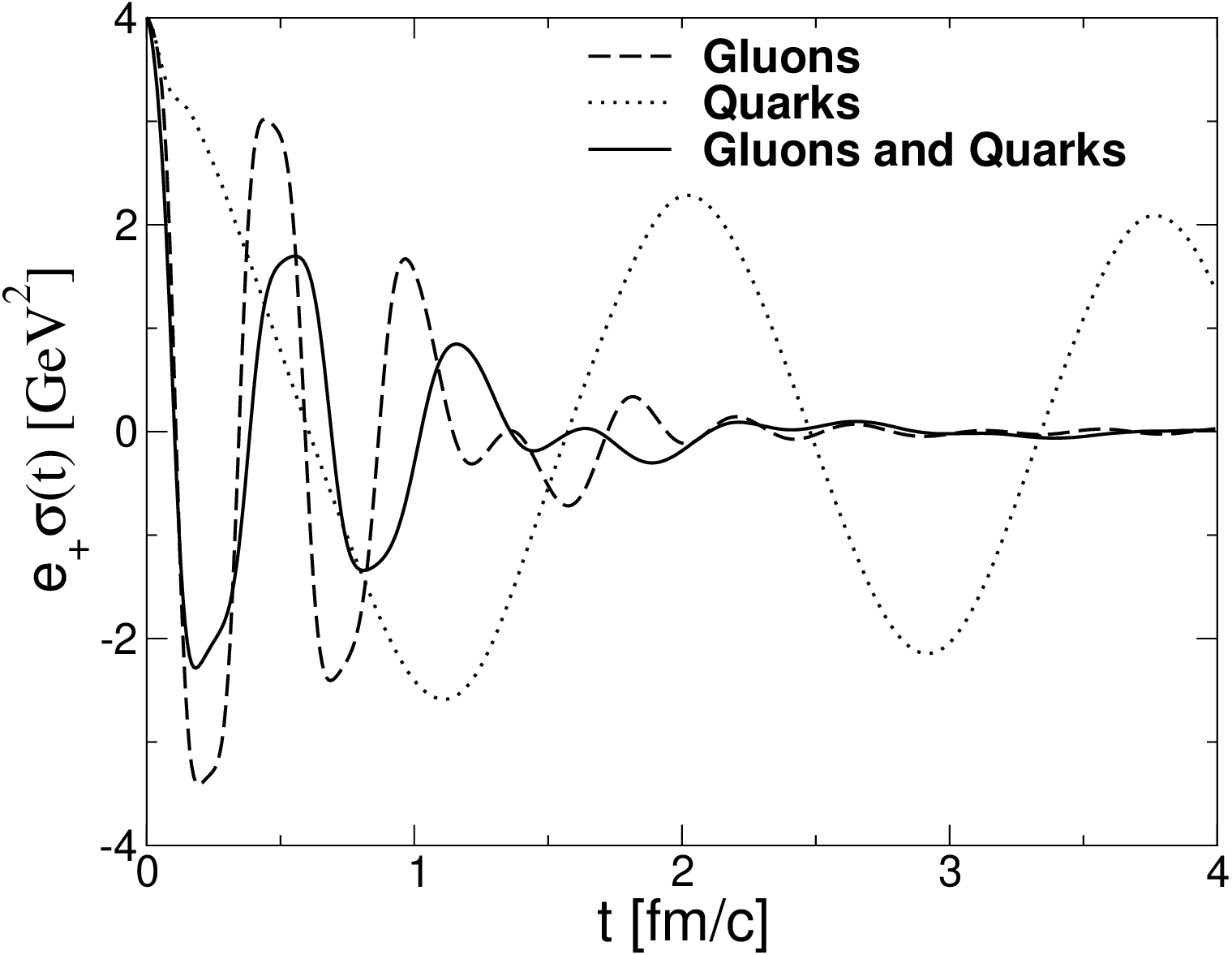,width=8.2cm,angle=0}}
\caption{The time dependence of the strong background ``chromo-electric'' field,
$\sigma(t)$,  for a
flux tube radius of $\rho_0 = 2$ fm for 
``quarks
and gluons''  with self-interaction (solid line), 
``gluons'' only with self-interaction (dashed line) and ``quarks'' only
(dotted line). The ``quarks'' are much less effected by the dynamical
coupling compared to the ``gluons'', {\it cf.} Fig. \ref{fig4}. Inclusion of the ``quarks'' leads to a
faster damping of the oscillations in the beginning of the evolution.
}
\label{fig3}
\end{figure}
The field energy is transformed into the energy of the produced partons,
i.e. the amplitude of the oscillations for $t < \tau_f$ is decreasing
with time. 
In the vicinity of large magnitudes of $\sigma(t)$
the current is zero since the partons stop their collective motion
within the flux tube and at this time the distribution function is
symmetrically  peaked around zero momentum \cite{vinnik}. For fermions
Pauli blocking is most efficient and no particles can be produced in already
occupied states. For bosons, however, the enhancement of particle
production has its maximum at this time and a large number of bosons
is produced. Therefore a direct flow of energy from the field to the
gluons occurs, i.e. the damping of the mean field happens very
sufficiently. With other words, the quantum statistical nature leads to
a fast energy transformation for bosons but to a suppressed one for fermions.

\begin{figure}[t]
\centerline{\epsfig{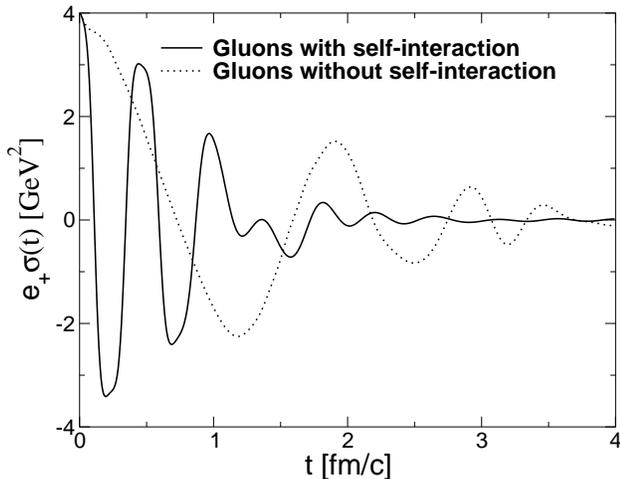}}
\caption{The time dependence of the strong background ``chromo-electric'' field,
$\sigma(t)$,  for ``gluons'' with (solid line) and
without (dotted line) self-interaction. The flux tube radius is
$\rho_0 = 2$ fm and $m_+= 0$. It is apparent that the inclusion of
self-interaction leads to a rapid collapse of the mean background
field. Without self-interaction the energy is smoothly converted from
the field to the particles. 
}
\label{fig4}
\end{figure}

The origin of the collapse of the flux tube is the back reaction
phenomenon \cite{motola}, the time scale on which this appears is given by
self-interaction, see Fig. \ref{fig4}. 
This becomes clear by neglecting the non-linear terms in
Eq. (\ref{motionmax}). That is one new result of the present investigation.
The
plasma oscillations are smoothly damped only,
i.e. the energy is slowly converted from the mean field to the
produced particles. In that case the background vanishes at much
larger times. 

There is only a weak  dependence of the results for the mean-field on the flux tube
radius. Using the different radii given in Table I, the time scale
$\tau_f$ is changed 
by about 10 \% but all
qualitative effects are unchanged. For values of $\Lambda^{IR} \sim
m_+$, the lower bound in momentum acts like a mass scale itself and
leads to similar results as discussed above in connection with the
finite gluon mass. Numerical examples of that consideration are given
in the next subsection.
\subsection{Particle production}
\begin{figure}[tb]
\centerline{\epsfig{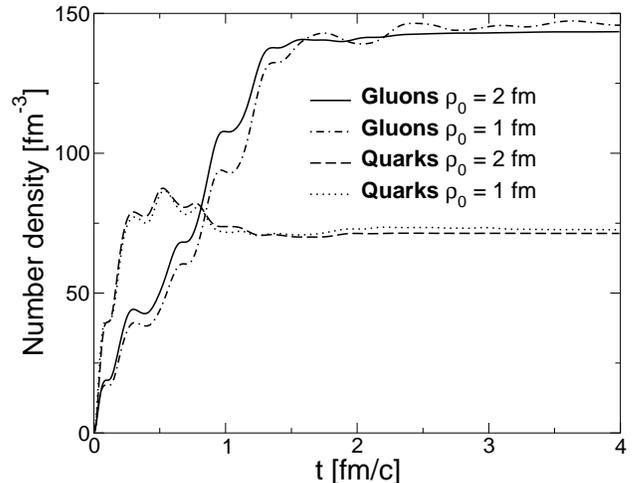}}
\caption{The time dependence of the number density for ``quarks and
gluons''. A fast increase of the particle number is observed at the
beginning of the evolution. At $\tau_f\sim 1.5$ fm/c the background
mean-field vanishes and therefore the creation process stops and the
number density is constant. 
Different flux tube radii do not alter the results significantly. 
}
\label{fig5}
\end{figure}

The self-consistent solution of the quantum kinetic equations for
quarks and gluons provides particle creation and a strong background
mean-field discussed in the 
previous section. The large strength
of the field initiates the particle production process and ``quark''- and
``gluon'' pairs are spontaneously produced. In Fig. \ref{fig5} we plot the
number density 
\begin{equation}\label{density}
n_\pm(t)=\frac{g_\pm}{\pi\rho_0^2}\sum_{nl}\int\frac{dp_\parallel}{2\pi}\,\,f_\pm(\vec
p,t)
\end{equation}
of the produced ``quarks and gluons''. It is simple to verify  that Eq. (\ref{density})
is a finite expression, e.g. by using the equations given in
Appendix B. At the beginning of the evolution the background field is
very large and therefore many particles are produced in a short
time. At the same time the background field decreases, energy is
converted from the field into the created particles. At  $\tau_f\sim 1.5$ fm/c the background
mean-field vanishes and consequently no more quarks and gluons are
produced and the number density assumes a constant value. The number
density is very large and therefore it is necessary to solve the full
non-Markovian equation (\ref{KE}). The application of the low density
limit would underestimate the production of ``gluons'' and overestimate
the ``quark'' production. Note that, in a
realistic description of the further evolution of the produced plasma,
collisions between the partons could equilibrate the
system. Additionally, the number
density would   decrease when an expansion is implemented.

The regularized energy density is defined by
\begin{equation}\label{energy}
\epsilon_\pm(t)=\frac{g_\pm}{\pi\rho_0^2}\sum_{nl}\int\frac{dp_\parallel}{2\pi}\omega_\pm(\vec
p)\,\,(f_\pm(\bar p,t)-f_\pm^c(\bar p,t))
\end{equation}
with the  counter term
\cite{Back,vinnik} 
\begin{equation}\label{counter}
f_\pm^c(\bar p,t) =
\bigg[\frac{e_\pm\sigma(t)p_\parallel}{4\omega^3_\pm(\bar p)}\bigg(
\frac{\varepsilon_{\pm \perp}}{p_\parallel}\bigg)^{2s_\pm}\bigg]^2\,.
\end{equation}
The time dependence of the energy density is plotted in
Fig. \ref{fig6}.
Due to the strong background mean-field many particles are produced at
small times and energy is converted from the field to the particles
(see Fig. \ref{fig2}). During the fast
increase of the energy density a shoulder-like structure appears. The
period of these waves is in tune with the oscillations of the mean
field.
The  energy density reached at large times for ``gluons'' is larger compared to the
``quarks'' due to Pauli blocking acting on the quarks. The value of about
$200$ GeV/fm$^3$ corresponds to a (quasi-equilibrium) temperature of about $1$ GeV and is
characteristic for planned ultra-relativistic heavy-ion collision
experiments.
The insertion in
Fig. \ref{fig6} shows the sensitivity to the ``gluon'' mass. A finite
gluon mass leads to weak plasma oscillations in the background
mean-field even for large times as discussed in connection with Fig. \ref{fig2}
and therefore the energy density oscillates in concert with these
repeated creation and annihilation processes. 

Naturally the question arises whether the system is in  equilibrium
after the production of particles has stopped and no fields are
imposed. Although the number- and energy density is constant for large
times,
the system is still in a strong non-equilibrium state. Further
interactions such as inter-parton collisions could provide
thermalization, but that is a challenging theoretical question on its own.
Further understanding can be provided by studying the pressure for the
``quarks and gluons'' given by the following expressions:
\begin{figure}[t]
\centerline{\epsfig{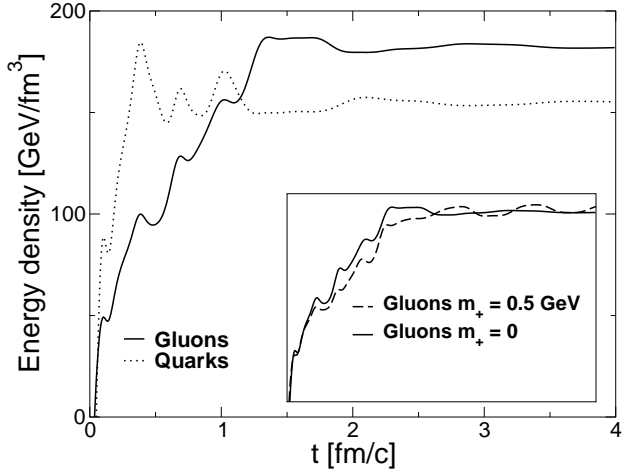}}
\caption{The time dependence of the energy density for ``quarks and
gluons'' for $\rho_0 = 2$ fm. Due to the collapse of the 
background mean
field, all energy is converted at $\tau_f\sim 1.5$ fm/c into the
partons and a constant value of the energy densities is assumed. 
Different ``gluon'' masses do not alter the results significantly, see insertion. 
}
\label{fig6}
\end{figure}
\begin{eqnarray}\nonumber
P_+^\parallel(t)&=&\frac{g_+}{\pi\rho_0^2}\sum_{nl}\int\frac{dp_\parallel}{2\pi}\bigg[\frac{p_\parallel^2}{\omega_+(\vec
p)}\bigg(f_+(\bar p,t) - f^c_+(\vec
p,t)\bigg)\\
&&+\bigg(\frac{p_\parallel^2}{2\omega_+(\bar p)}
-\frac{\omega_+(\vec
p)}{6}\bigg)\bigg(v_+(\bar p,t) - v^c_+(\vec
p,t)\bigg)\bigg],\\\nonumber
P_+^\perp(t)&=&\frac{g_+}{2\pi\rho_0^2}\sum_{nl}\int\frac{dp_\parallel}{2\pi}\bigg[\frac{(\varepsilon^+_{nl})^2}{\omega_+(\vec
p)}\bigg(f_+(\bar p,t) - f^c_+(\vec
p,t)\bigg)\\
&&+\bigg(\frac{(\varepsilon^+_{nl})^2}{2\omega_+(\bar p)}
-\frac{\omega_+(\vec
p)}{3}\bigg)\bigg(v_+(\bar p,t) - v^c_+(\vec
p,t)\bigg)\bigg],\\\nonumber
P_-^\parallel(t)&=&\frac{g_-}{\pi\rho_0^2}\sum_{nl}\int\frac{dp_\parallel}{2\pi}
\bigg[\frac{p_\parallel^2}{\omega_-(\bar p)}\bigg(f_-(\bar p,t) - f^c_-(\bar p,t)\bigg)\\
&&+\frac{\varepsilon^-_\perp p_\parallel}{2\omega_-}\bigg(v_-(\bar p,t) - v^c_-(\vec
p,t)\bigg)\bigg],\\\nonumber
P_-^\perp(t)&=&\frac{g_-}{2\pi\rho_0^2}\sum_{nl}\int\frac{dp_\parallel}{2\pi}
\bigg[\frac{(\varepsilon^-_{nl})^2}{\omega_-(\bar p)}\bigg(f_-(\bar p,t) - f^c_-(\bar p,t)\bigg)\\
&&-\frac{(\varepsilon^-_{nl})^2 p_\parallel}{2\omega_-(\bar p)\varepsilon_\perp}\bigg(v_-(\bar p,t) - v^c_-(\vec
p,t)\bigg)\bigg],
\end{eqnarray}
where the regularizing counter term $f^c$ is given in
Eq. (\ref{counter}) and $v^c$ reads
\begin{equation}
v_\pm^c(\bar p,t) = \frac{e_\pm\dot \sigma(t)p_\parallel}{8\omega^4_\pm(\bar p)}\bigg( \frac{\varepsilon_{\pm\perp}}{p_\parallel}\bigg)^{2s_\pm}\,.
\end{equation}
In Fig. \ref{fig7} we plot the different components of the pressure.
 It is apparent that the considered system is
still out-off equilibrium. It is interesting to observe that the
difference is much more pronounced for ``gluons'' (lower panel) compared
to ``quarks'' (upper panel). The ``gluon'' production is
unhindered and preferable in longitudinal direction, for ``quarks'' again
Pauli blocking prevents a drastic difference.
In an equilibrated system the longitudinal and the parallel pressure
contributions  would be equal.
\begin{figure}[t]
\centerline{\epsfig{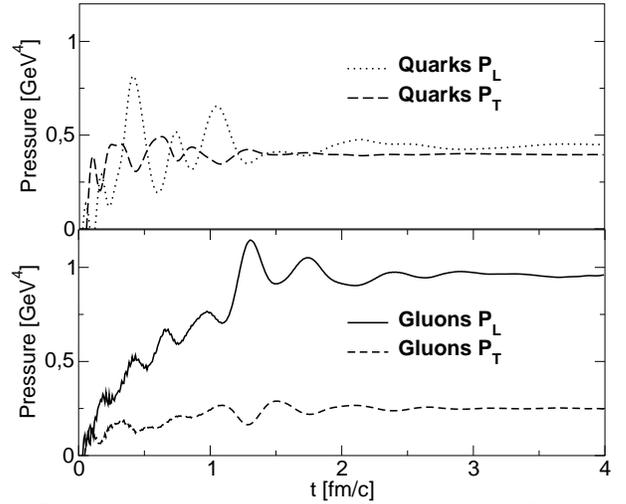}}
\caption{The time dependence of the pressure of ``gluons'' (lower panel) and
``quarks'' (upper panel). The difference between the longitudinal and transverse part
indicates that the system is out-off equilibrium also at large
times. Pauli blocking prevents that the difference is less pronounced
for ``quarks'' compared to ``gluons''.
}
\label{fig7}
\end{figure}
%

\section{Summary}
We have described coupled fermion and boson production within a
quantum kinetic approach in cylinder geometry. The strong background
field is given by a
dynamical mean-field containing self-interactions. Back reactions are
included, both fermionic and bosonic currents modify the initial
``chromo-electric'' background field. Strong particle creation appears at the
beginning of the time evolution, however due to back reactions the
background mean-field vanishes. The time scale at which such a collapse
of the flux tube happens depends on the self-interaction. This general
approach has been applied to the production of "quarks" and ``gluons''
and using a
zero ``gluon'' mass, a momentum dependent quark mass from Dyson-Schwinger
studies in concert with a flux tube radius of about $2$ fm, we find a
typical time scale of $1.5$ fm/c for this effect. 
At that time all energy is converted from the field to the
particles and the evaluated number- and energy  densities reach a
constant value being of the typical order of magnitude of a few
hundred GeV/fm$^3$. The system is still out off equilibrium, we have
exemplified this by calculating the pressure components which are
different in longitudinal and transverse direction.

The further evolution of the system will strongly depend on
additionally included interactions, such as collisions. A detailed
study will be reported elsewhere.

\section*{Acknowledgment} 
We thank C.D. Roberts  for helpful discussions.
This work was supported by Deutsche
Forschungsgemeinschaft under project number SCHM 1342/3-1, AL 279/3-3
and 436~RUS~17/102/00;  and partly by the
Russian Federation's State Committee for Higher Education under grant 
N E00-33-20.

\appendix
\section{Vector interaction and bosons}
In Section II we have introduced an effective Lagrangian in mean field
approximation based on coupled boson and fermion fields. The boson
fields have been considered as scalar fields. In this appendix we will
briefly demonstrate that starting from a charged vector field, the
same Lagrangian in mean field approximation can be obtained.

The fermionic contribution is identical to Eq. (\ref{lagrferm}) and
therefore restrict ourselves to the bosonic contribution. 
The Lagrangian for a minimal substituted  charged vector field reads:
\begin{eqnarray}\label{100}
L^{(0)} (x) =  \left( D_\nu  \omega_\mu (x) \right)^* 
\left( D^\nu  \omega^\mu (x) \right)
-  m_+ ^2 \omega_\mu^*\omega^\mu\,,
\end{eqnarray}
where $D_\nu=\partial_\nu+ie_g\omega_\nu$  is the  covariant derivative.
The decomposition into a mean-field contribution and fluctuations reads
\begin{eqnarray}\label{110}
\omega_\mu=\Omega_\mu+\delta_\mu, \qquad \left< \omega_\mu \right>=\left< \Omega_\mu \right>\,.
\end{eqnarray}
Using this Ansatz in Eq. (\ref{100}) we obtain
\begin{eqnarray}
\label{120}
&& \left( D_\nu  \omega_\mu (x) \right)^* \left( D^\nu  \omega^\mu (x) \right) =\nonumber\\
&& \left( \partial_\nu \Omega_\mu^* \partial^\nu \Omega^\mu +
         \partial_\nu \delta_\mu^* \partial^\nu \delta^\mu -
         e_+^2 \left| \Omega \right|^4 -
         e_+^2 \left| \Omega \right|^2 \left| \delta \right|^2  \right) \nonumber \\
&&+ie_+ \left[ \partial_\nu \delta_\mu^* \delta^\nu \Omega^\mu - \delta_\nu^* \Omega_\mu^* \partial^\nu \delta^\mu \right]\nonumber\\
&&+ie_+ \left[ \partial_\nu \Omega_\mu^* \delta^\nu \delta^\mu - \delta_\nu^* \delta_\mu^* \partial^\nu \Omega^\mu \right]\,,
\end{eqnarray}
where we have neglected odd orders in the fluctuations
$\delta$. Furthermore we focus on the collisionless limit, i.e. we do
not consider  $e_+^2 \left| \delta \right|^4$ terms.

Assuming that the fluctuations do not depend on the mean-field part and
imposing the initial spatial anisotropy of the flux tube, we can write
in Hartree approximation
\begin{eqnarray}\label{130}
\left< \delta_k^* \delta^k \right>=3 \left< | \varphi |^2 \right>\,, \qquad
\left< \delta_i^* \delta_k \right>=\delta_{ik} \left< | \varphi |^2 \right> \,,
\end{eqnarray}
where $i,k=1,2,3$. $\varphi$ is the effective fluctuating scalar
field. The mean-field reads
\begin{eqnarray}\label{140}
\Omega_\mu=\left( 0,0,0,\Phi_0(t) \right)\,.
\end{eqnarray}
Employing this Hartree-Fock like approximation in 
temporal gauge we can construct the following modified Lagrangian
\begin{eqnarray}
\label{150}
L_{\varphi}&=& 3\partial_\nu \varphi^* \partial^\nu \varphi - 3m_+^2  |\varphi |^2
-3e_+^2 \left| \varphi \right|^2  \Phi_0^2(t) \nonumber\\
&&+ ie_+ \Phi_0(t) \left[ \partial_3 \varphi^* \varphi - \varphi^* \partial_3 \varphi \right]\,.
\end{eqnarray}
Eq. (\ref{150}) can be generalized, using
$\partial_3\longrightarrow\nabla_n$, to obtain
the Lagrangian Eq. (\ref{lagrbos}).
\section{Equations for the numerical solution}
Eq. (\ref{KE}) is an integro-differential equation. It can be
re-expressed by introducing
\begin{eqnarray}\nonumber
u_\pm(\vec{p},t) &\equiv& \int_0^t dt^{\prime}
W_\pm(\vp,t,t')
\Big( 1 \pm 2f_\pm(P_\pm(t,t'),t') \Big)\times\\
&& \sin[2\int_{t'}^t d\tau \omega_\pm(t,\tau)],
\label{eq: vosemcet}\\\nonumber
v_\pm(\vec{p},t) &\equiv& \int_0^t dt^{\prime}
W_\pm(\vp,t,t')
\Big( 1 \pm 2f_\pm(P_\pm(t,t'),t^{\prime}) \Big)\times\\
&& \cos[2\int_{t'}^t d\tau \omega_\pm(t,\tau)],
\label{eq: vosempet}
\end{eqnarray}
with the initial conditions $u(\vec{p},0)=v(\vec{p},0)=0$, in which case
we find
\begin{eqnarray}
\frac{{\partial f_\pm (\vp ,t)}} {{\partial t}} +
e_\pm\sigma(t)\frac{{\partial f_\pm (\vp ,t)}} {{\partial p_\parallel }}& =&
\frac{1}{2}W_\pm(\vp,t) v_\pm(\vec{p},t),
\label{eq: vosemshest}\\
\frac{{\partial v_\pm (\vp ,t)}} {{\partial t}} +
e_\pm\sigma(t)\frac{{\partial v_\pm (\vp ,t)}} {{\partial p_\parallel }}& =& W_\pm(\vp,t)\Big( 1 \pm 2f_\pm(\vec{p},t) \Big)\nonumber\\
&-& 2\omega_\pm(\vp) u_\pm(\vec{p},t),
\label{eq: vosemsem}\\
\frac{{\partial u_\pm (\vp ,t)}} {{\partial t}} +
e_\pm\sigma(t)\frac{{\partial u_\pm (\vp ,t)}} {{\partial p_\parallel }}& =& 2{\omega}_\pm(\vp) v_\pm(\vec{p},t).
\label{eq: vosemvosem}
\end{eqnarray}
where the total energy is defined in Eq. (\ref{350}) for bosons and in
Eq. (\ref{190}) for fermions. The transition amplitudes are given by
\begin{eqnarray}
 W_\pm(\vp,t)&=&\frac{e_\pm\sigma(t)p_\parallel}{\omega^2_\pm(\vp)}\bigg(\frac{\varepsilon _{\pm \bot nl }}{p_\parallel}\bigg)^{2s_\pm}\,.
\end{eqnarray}
For the currents we obtain the following final expression
\begin{eqnarray}\label{360}
j_\pm(t)&=&2e_\pm g_\pm \frac{1}{\pi \rho_0^2} \sum_{nl} \int \frac{dp_\parallel}{2\pi} 
\frac{p_\parallel}{\omega_\pm(\vp)}\,\bigg(
f_\pm(\vec{p},t)\\\ \nonumber
&+&\frac{1}{2} \,v_\pm(\vec{p},t)
\bigg[\frac{\epsilon_{\bot nl}}{p_\parallel}\bigg]^{2s_\pm}
-e_\pm\dot \sigma(t)
\frac{p_\parallel}{8\,\omega_\pm^{4}(\vp)}\bigg[\frac{\epsilon_{\bot nl}}{p_\parallel}\bigg]^{2s_\pm}\bigg)\,.
\end{eqnarray}
In concert with Eq. (\ref{motionmax}) they define the mean background
field $\Phi_0(t)$.


\end{document}